\documentclass[letterpaper, 10 pt, conference]{ieeeconf}   

\IEEEoverridecommandlockouts                              

\usepackage{graphics} 
\usepackage{verbatim}
\usepackage{epsfig} 
\usepackage{times} 
\usepackage{amsmath} 
\usepackage{amssymb}  
\usepackage{physics}
\usepackage{makecell}
\usepackage{xcolor}
\title{\LARGE \bf
Consensus control with safety guarantee: an application to the kinematic bicycle model
}
\author{K. Niu$^{1}$, C.T. Abdallah$^{2}$, and M. Hayajneh$^{3}$
\thanks{$^{1}$School of Electrical and Computer Engineering, Georgia Institute of Technology, Atlanta, GA 30332, USA.
        {\tt\small kniu9@gatech.edu}}%
\thanks{$^{2}$School of Electrical and Computer Engineering, Georgia Institute of Technology, Atlanta, GA 30332, USA.
        {\tt\small ctabdallah@gatech.edu}}
\thanks{$^{3}$College of Information Technology, United Arab Emirates University, Sheik Khalifa Bin Zayed Street, P. O. Box 15551, Al-Ain, UAE.
		{\tt\small mhayajneh@uaeu.ac.ae}}%
}

\begin{document}

\maketitle
\thispagestyle{empty}
\pagestyle{empty}

\begin{abstract}
This paper proposes a consensus controller for multi-agent systems that can guarantee the agents' safety. The controller, built with the idea of output prediction and the Newton-Raphson method, achieves consensus for a class of heterogeneous nonlinear systems. The Integral Control Barrier Function is applied in conjunction with the controller, such that the agents' states are confined within pre-defined safety sets. Due to the dynamically-defined control input, the resulting optimization problem from the barrier function is always a Quadratic Program, despite the nonlinearities that the system dynamics may have. We verify the proposed controller using a platoon of autonomous vehicles modeled by kinematic bicycles. A convergence analysis of the leader-follower consensus under the path graph topology is conducted. Simulation results show that the vehicles achieve consensus while keeping safe inter-agent distances, suggesting a potential in future applications.

\end{abstract}

\section{Introduction}
Consensus control has been extensively investigated in the setting of multi-agent systems, where typically it is underscored by a distributed algorithm that guarantees convergence of the state or output variables of various agents to a common target value; see, e.g., \cite{mesbahi2010graph} and references therein. Application areas of consensus control include swarms of unmanned aerial vehicles and platoons of self-driving cars. Following the initial results for homogeneous linear systems, recent studies focused on consensus for nonlinear heterogeneous systems. Ref. \cite{ren2009distributed} derived a consensus controller for heterogeneous Euler-Lagrange systems; \cite{ding2013consensus} investigated general nonlinear single-input-single-output systems; \cite{yin2019second} considered a class of second-order systems. Consensus controllers using  adaptive control  \cite{feng2021adaptive} \cite{an2021decentralized} and Model Predictive Control (MPC)  \cite{xiao2019leader} \cite{gao2017distributed} are also proposed.%

Recently, the authors of this paper developed a consensus controller for heterogeneous nonlinear systems \cite{niu2023consensus} using the agents' predicted outputs after a given time horizon. With the Newton-Raphson method, consensus may be achieved for a class of nonlinear systems. However, the controller does not provide any measure of safety. For example, it allows multiple agents (e.g., autonomous vehicles) to occupy common physical spaces as the consensus control converges, leading to inter-agent collisions that are unfavorable in practice. To satisfy safety constraints, this paper supplements a controller with a Control Barrier Function (CBF), such that the states of the agents are restricted within pre-defined safety sets. Thanks to the dynamically-defined control inputs, the Integral Control Barrier Function (I-CBF) \cite{ames2020integral} as a special form of the CBF may be used. Despite the nonlinear dynamics of the system, the optimization problems associated with the I-CBF are always Quadratic Programs (QPs), which is different from the classical CBF that may result in Nonlinear Programs (NLPs). This feature enables a faster solution and an easier feasibility check during the control period. While it is also possible to maintain safety by directly modifying the consensus controller, a CBF technique may serve as a backup to unforeseen circumstances and would modify the controller in a less aggressive manner. %

As a practical application, this paper considers a platoon of autonomous vehicles. Control of autonomous vehicles has been extensively studied in the literature. Adaptive Cruise Control focuses on vehicle longitudinal dynamics, for which a PID controller is designed in \cite{milanes2014cooperative}, and an MPC controller in \cite{bageshwar2004model}. The lateral dynamics, which are nonlinear and can be described by kinematic or dynamic bicycle models \cite{kong2015kinematic}, have been studied as well. Ref.  \cite{wu2018consensus} designed a controller for vehicle merging and platooning. A hierarchical formation controller for platoons has been derived in \cite{qian2016hierarchical} using MPC. Ref.   \cite{khalifa2018vehicle} proposes a tracking controller for the lateral dynamics. The CBF technique is also applied to autonomous vehicles. \cite{ames2014control} applied CBFs for Adaptive Cruise Control. \cite{ames2019control} achieved lane keeping and spacing on unicycle models. \cite{seo2022safety} applied the CBF with a tracking technique to the dynamic bicycle. While many existing results study the vehicle's lateral and longitudinal dynamics separately, this paper considers them jointly to demonstrate the interactions between platooning and CBFs in a 2-D plane. We also recognize that in practice, it is often desired that the vehicle platoon could follow a pre-defined trajectory. Therefore, this paper considers the consensus of the leader-follower structure, as an extension of the leaderless controller \cite{niu2023consensus}.
A convergence analysis is performed for the platoon modeled by directed path graphs. We evaluate the safety of the platoon by the inter-agent distances. Using the I-CBF, each agent maintains a proper distance to its successor and predecessor, avoiding potential collisions in unexpected conditions. %

The remainder of this paper is organized as follows: Section II  reviews the existing consensus controller and summarizes relevant results on I-CBFs. Section III defines the consensus problem for the vehicle platoons and presents the proposed solutions. Section IV discusses the simulation results, and Section V concludes the paper.   %

\section{Survy of background material}
\label{sec_prelim}
\subsection{Prediction-based consensus controller}
The consensus controller \cite{niu2023consensus} features the idea of the output prediction and the Newton-Raphson method. This approach was inspired by \cite{Arxiv}, where a tracking controller for single-agent systems is derived. Consider a multi-agent system consisting $N$ agents, denoted by $A_{i}$,    $i\in \mathcal{N}:= \{1,2,3,\hdots, N\}$. Suppose that the motion dynamics of $A_{i}$ are modelled by the differential equation:
\begin{equation}
\label{eqn_mas_dynamics}
    \dot{x}_i(t) = f_i(x_i(t),u_i(t)),
\end{equation}
where $x_i(t) \in \mathbb{R}^{n_i}$ and $u_i(t) \in \mathbb{R}^{m}$ are the respective state variable  and input variable  of $A_i$, for some given positive integers $n_{i}$ and $m$. Denote 
$x_{i}(0)\in\mathbb{R}^{n_{i}}$ as the initial condition of Eqn. (\ref{eqn_mas_dynamics}) at time $t=0$. Suppose that 
$f_i : \mathbb{R}^{n_i}\times \mathbb{R}^m \mapsto \mathbb{R}^{n_i}$ is a continuously differentiable function. In addition, $f_i$ satisfies suitable sufficient conditions for the existence of unique solutions of Eqn. (1), in the time interval $t\in[0,\infty)$ for every bounded, piecewise continuous  input signal $\{u_i(t):t\in[0,\infty)\}$, and initial condition $x_i(0)\in\mathbb{R}^{n_{i}}$ (see, e.g., Ref. \cite{niu2023consensus}, Assumption 1). The output of the agent $A_i$ is denoted by
\begin{equation}
\label{eqn_MAS_outputs}
    y_i(t) = h_i(x_i(t)),
\end{equation}
where the function $h_i : \mathbb{R}^{n_i}\mapsto\mathbb{R}^m$ is continuously differentiable. Observe that the dimensions of the agents' state spaces, $n_{i}$, may be different from each other, but their input and output spaces must have the same dimension, $m$. %

The consensus problem requires a distributed controller, such that the outputs of all agents can (asymptotically) converge to the same point, namely:
\begin{equation}
\label{eqn_consensus_requirement}
    \limsup_{t \rightarrow \infty} \Vert y_i(t) - y_j(t)\Vert = 0, \forall i, j \in \mathcal{N}.
\end{equation} %

To design such a controller, information exchange among agents is often necessary. This information exchange is carried out over a communication network, whose topology is modeled by an undirected graph $\mathcal{G}$. Its vertices, $V_{i}$, $i=1,\ldots,N$, correspond to the respective agents $A_{i}$, $i=1,\ldots,N$. Its edges, connecting pairs of vertices, correspond to bi-directional communication links between two agents.
$A_{j}$ is a neighbor of $A_{i}$ if and only if $A_{i}$ can receive information from $A_{j}$. The indices of the neighbors of the $i$-th agent are denoted by a set $N_i$. For the leaderless consensus controller over undirected graphs, we make the following assumption:

\textit{Assumption I}: The (undirected) communication graph $\mathcal{G}$ is connected, meaning that there is a path formed by a sequence of neighboring edges between every pair of vertices $(V_i, V_j), i\not=j$. \hfill $\Box$%

The consensus controller \cite{niu2023consensus} attempts to reach a consensus among the agents' predicted outputs. Suppose that all agents have the same prediction horizon, $T>0$, and denote the predicted output of $A_i$, computed at time $t$, by $\tilde{y}_i(t+T)$. We make the following assumption:

\textit{Assumption II}: The predicted output $\tilde{y}_i(t+T)$ functionally depends on $(x_{i}(t)^\top,u_{i}(t)^\top)^\top$ in the following way:
\begin{equation}
\label{eqn_predictor}
    \tilde{y}_i(t+T) = g_i(x_i(t), u_i(t)),
\end{equation}
where $g_i : \mathbb{R}^{n_i} \times \mathbb{R}^m \mapsto \mathbb{R}^{m}$ is a continuously differentiable function.\hfill  $\Box$

We point out that the function $g_i(\cdot, \cdot)$ may not have a known analytic form, but it can be approximated by numerical methods or calculated by simulations.

The consensus controller \cite{niu2023consensus} has the following form:
 \begin{equation}
 \label{eqn_consensus_controller}
     \dot{u}_{i}=-\alpha_{i}\left(\frac{\partial g_{i}}{\partial u_i}(x_{i},u_{i})\right)^{-1}\sum_{j\in N_{i}}\big(g_{i}(x_{i},u_{i})-g_{j}(x_{j},u_{j})\big),
 \end{equation}
where $\alpha_i \geq 1$ is a given constant called the controller's speedup factor of $A_{i}$. As discussed in \cite{niu2023consensus} and explained in the sequel, large controller speedup factors may be associated with the stabilization of the closed-loop system and reductions of asymptotic tracking errors. Note that Eqn. (\ref{eqn_consensus_controller}) is a fluid-flow version of the Newton-Raphson method, which takes the average of the solutions $u_i$ of $g_i(\cdot) = g_j(\cdot), j \in N_i$. Under a bounded-input-bounded-state stability condition defined in \cite{niu2023consensus}, there exists a positibe number $\eta$, such that the local consensus error is bounded by:
 \begin{equation}
\lim\sup_{t\rightarrow\infty}||\sum_{j\in{\mathcal N}_{i}}\big(\tilde{y}_{i}(t+T)-\tilde{y}_{j}(t+T)\big)||^2<\frac{\eta}{4\alpha_{\min}},
 \end{equation}
where $\alpha_{{\rm min}}:={\rm min}\{\alpha_{i}:i\in{\mathcal N}\}$. Moreover, as $\alpha_{\min}\rightarrow \infty$, the global consensus error converges to $0$. Note that the controller achieves consensus on the agents' predicted outputs instead of true outputs. The difference between predictions and true outputs, called the prediction gap, can only be reduced by using high-precision predictors $g_i(\cdot)$. %

\subsection{Integral Control Barrier Function}
The Control Barrier Function (CBF) technique is an effective method that can ensure the safety of feedback control systems. The objective of the CBF is to confine the system states within a pre-defined admissible set while modifying the system inputs in the least invasive way. As a specific form of the CBF, the Integral Control Barrier Function (I-CBF) (\cite{ames2020integral}) is dedicated to the dynamically-defined controllers. Consider a dynamically controlled system:
\begin{equation}
    \label{eqn_dynamic_control}
    \begin{split}
    \dot{x}(t) &= f(x(t),u(t))\\
    \dot{u}(t) &= \Phi(x(t), u(t)),
    \end{split}
\end{equation}
where $f:\mathbb{R}^{n}\times \mathbb{R}^{m}\rightarrow \mathbb{R}^{n}$ is the state equation of the system, and $x(t) \in \mathbb{R}^n$ are the system states. The inputs $u(t)\in\mathbb{R}^m$ of the system are dynamically defined by a continuous function $\Phi:\mathbb{R}^n\times \mathbb{R}^m\rightarrow\mathbb{R}^{m}$. This setting is different from a static controller $u(t) = \phi(x(t))$. Let $S\subset \mathbb{R}^n \times \mathbb{R}^m$ be a closed set (called the safety set), such that the system is safe if and only if $z(t) \triangleq (x(t)^\top, u(t)^\top)^\top\in S$\footnote{Compared to the traditional CBF, the definition of the safety set $S$ has been extended to encompass the system inputs $u(t)$. This is due to the fact that the controller is dynamically defined. See \cite{ames2020integral} for a detailed discussion.}.  Safety control requires a sequence of control input $u(t), t \in [0, +\infty)$ for (\ref{eqn_dynamic_control}), so that the set $S$ is:
\begin{enumerate}
    \item forward-invariant, meaning that if $z(t_0) \in S$, then $z(\tau) \in S$ for all $ \tau\in[t_0,+\infty)$.
    \item  exponentially stable,  meaning that the distance from $z(t), t \in [0,+\infty)$ to $S$ reduces exponentially. 
\end{enumerate}
To ensure these safety requirements, let $h:\mathbb{R}^n \times \mathbb{R}^m \rightarrow \mathbb{R}$ be a continuously-differentiable function satisfying the following condition:
    \begin{equation}
    \begin{split}
        &h(z(t)) > 0,~~~~~ \forall z(t)\in\mathcal{S} - \partial\mathcal{S}\\
        &h(z(t)) = 0,~~~~~ \forall z(t)\in\partial
        \mathcal{S}\\
        &h(z(t)) < 0,~~~~~ \forall z(t)\not\in\mathcal{S}.
    \end{split}
\end{equation}
Let $\kappa:\mathbb{R}\rightarrow\mathbb{R}$ be an extended class-$\mathcal{K}$ function. Suppose that for every system trajectory $\{z(t):t\in[0,\infty)\}$ the following equation is satisfied for every $t\in[0,\infty)$:
\begin{equation}
    \label{eqn_dh_safety}
    \dot{h}(z(t)) + \kappa(h(z(t))) \geq 0,
\end{equation}
then the forward invariance and the exponential stability of the set $S$ can be guaranteed (\cite{ames2014control,ames2019control}). In case that the original control law $\Phi(x(t), u(t))$ does not satisfy these requirements, consider modifying the dynamics of $\dot{u}(t)$ by adding a bias term $w(t) \in \mathbb{R}^{m}$:
\begin{equation}
    \label{eqn_dynamic_control_icbf}
    \begin{split}
    \dot{x}(t) &= f(x(t),u(t))\\
    \dot{u}(t) &= \Phi(x(t), u(t)) + w(t).
    \end{split}
\end{equation}
With the bias $w(t)$, an input $u(t)$ satisfying the safety constraint may be calculated. To avoid modifying the original control $\Phi(x(t), u(t))$ excessively, $w(t)$ should be as small as possible, leading to the following optimization problem:
\begin{equation}
\label{eqn_icbf_qp}
\begin{split}
    &\min_{w} \Vert w\Vert^2\\
    \text{s.t.}\ \  (\dfrac{\partial h}{\partial u}(x,u))^\top w \geq  &-(\dfrac{\partial h}{\partial x}(x,u))^\top f(x,u)\\- (\dfrac{\partial h}{\partial u}(x,u))^\top &\Phi(x, u) - \kappa(h(x,u)). 
\end{split}
\end{equation}
The inequality constraints in Eqn. (\ref{eqn_icbf_qp}) are direct results from Eqn. (\ref{eqn_dh_safety}).
The function $h(z(t))$ is a valid I-CBF if $\dfrac{\partial h}{\partial u} \not= 0$ or Eqn. (\ref{eqn_dh_safety}) holds when $\dfrac{\partial h}{\partial u} = 0$. If $h(\cdot)$ is invalid, higher-order control barrier function may be used. Define:
\begin{equation}
    h^{(2)}(z(t)) = \dot{h}(z(t)) + \kappa(h(z(t))). 
\end{equation}
If $\dfrac{\partial h^{(2)}}{\partial u} \not = 0$, then the bias $w$ can be calculated by:
\begin{equation}
\begin{split}
    &\min_{w} \Vert w\Vert^2\\
    \text{s.t.}\ \  (\dfrac{\partial h^{(2)}}{\partial u}(x,u))^\top w \geq  &-(\dfrac{\partial h^{(2)}}{\partial x}(x,u))^\top f(x,u)\\- (\dfrac{\partial h^{(2)}}{\partial u}(x,u))^\top &\Phi(x, u) - \kappa(h^{(2)}(x,u)),
\end{split}
\end{equation}
which also guarantees the safety of the set $\mathcal{S}$. If $h^{(2)}(\cdot)$ is still invalid, continue to construct higher order I-CBFs:
\begin{equation}
    \begin{split}
        h^{(3)}(z(t)) &= \dot{h}^{(2)}(z(t)) + \kappa(h^{(2)}(z(t))),\\
         h^{(4)}(z(t)) &= \dot{h}^{(3)}(z(t)) + \kappa(h^{(3)}(z(t))),\\
         &\hdots
    \end{split}
\end{equation}
until a valid one is found. By satisfying the higher-order I-CBFs, the safety of the set $\mathcal{S}$ can also be guaranteed. See \cite{ ames2020integral} for more details.%

We remark that since the I-CBF is native to the dynamically defined controller, it is natural to apply the I-CBF instead of the traditional CBF  to the consensus controller defined by (\ref{eqn_consensus_controller}) and (\ref{eqn_leader_follower_controller}) in the sequel. Plus, the optimization problem associated with the I-CBF is always a Quadratic Program, even if the controlled plant is nonlinear or non-control-affine. This property is one of the main advantages of our proposed approach.

\section{Leader-follower consensus of kinematic bicycles}
\subsection{The kinematic bicycle model}
\begin{figure}
    \centering
    \includegraphics[width = 3.5cm]{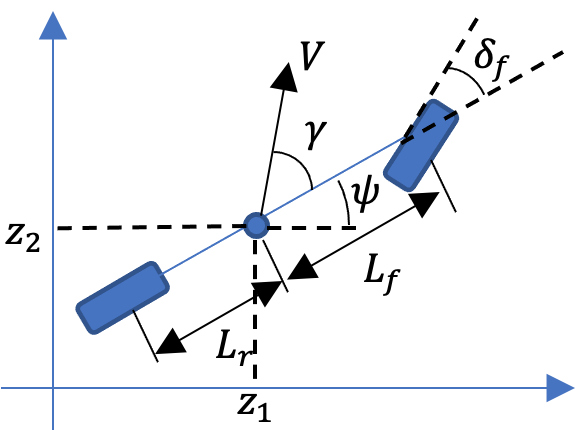}
    \caption{The kinematic bicycle model}
    \label{fig_bicycle_model}
    \vspace*{-2em}
\end{figure}

The kinematic bicycle model is a 4-th order nonlinear system, depicted in Fig. \ref{fig_bicycle_model}. The dynamics of this model are:
\begin{equation}
\label{eqn_bicycle_dynamics}
    \begin{split}
        &\dot{z}_1 = V\cos(\psi + \gamma)\\
        &\dot{z}_2 = V\sin(\psi + \gamma)\\
        &\dot{V} = a\\
        &\dot{\psi} = \dfrac{V}{L_r}\sin(\gamma),
    \end{split}
\end{equation}
where $(z_1, z_2)^\top \in \mathbb{R}^2$ represents the location of the bicycle in the 2-D plane. $V\in\mathbb{R}$ is the velocity of the bicycle. $a\in[a_{\min}, a_{\max}]$ is the bicycle's acceleration, and $\gamma \in [\gamma_{\min}, \gamma_{\max}]$ is the direction of the bicycle's velocity with respect to the heading of the bicycle frame. $L_r$ is the distance from the rear wheel to the bicycle's center of gravity (COG). Denote the angle of the steering wheel (with respect to the bicycle's heading) with $\delta_f \in (-\frac{\pi}{2},\frac{\pi}{2})$, then:
\begin{equation}
    \label{eqn_gamma_delta}
    \tan(\gamma) = \dfrac{L_r}{L_f+L_r}\tan(\delta_f),
\end{equation}
where $L_f$ is the distance from the front wheel to the bicycle's COG. The bijection (\ref{eqn_gamma_delta}) enables us to convert the input $\gamma$ to the physical control $\delta_f$ in reality. The system inputs are chosen to be $u = (a, \gamma)^\top$, and the system outputs are $y = (z_1, z_2)^\top$. To distinguish between different agents, we denote the states of agent $A_i$ by $x_i = (z_{i,1}, z_{i,2}, V_{i}, \psi_i)^\top$, the inputs of $A_i$ by $u_i = (a_i, \gamma_i)^\top$ and the outputs of $A_i$ by $y_i = (z_{i,1}, z_{i,2})^\top$. The distance between the COG and the rear wheel (and front wheel) of $A_i$ is denoted by $L_{i,r}$ (and $L_{i,f}$).%

\subsection{Leader-follower consensus for the vehicle platoon}
The controller proposed in \cite{niu2023consensus} considers only the case of leaderless consensus. As an extension to the original algorithm, this paper further studies the leader-follower consensus. Consider the multi-agent system consisting $K+1$ agents, denoted by $A_i, i \in \{0, 1, 2, \hdots, K\}$. The leader of the system, $A_0$, is autonomous, meaning that its control input $u_0$ is pre-defined rather than calculated by (\ref{eqn_consensus_controller}). In reality, this leader can be replaced by an external reference signal. The followers, $A_i, i = 1,2,3,\hdots,K$, are modeled by dynamics (\ref{eqn_mas_dynamics}) and outputs (\ref{eqn_MAS_outputs}). The leader-follower consensus requires that the output of the followers, $y_i(t), i = 1,2,3,\hdots,K$, could asymptotically converge to the output of the leader, $y_0(t)$, namely:
\begin{equation}
\label{eqn_leaderfollower_require}
    \lim\sup_{t\rightarrow\infty} \Vert y_i(t) - y_{0}(t)\Vert = 0, \forall i = 1,2,3,\hdots,K.
\end{equation}
The advantage of the leader-follower consensus over the leaderless consensus is that agents can track a given trajectory specified by the leader $A_0$ (or an external reference signal). This asymptotic tracking may sometimes be more preferable in the control of vehicle platoons.%
\begin{figure}[ht]
    \centering
    \includegraphics[width = 4cm]{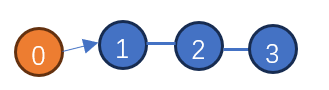}
    \caption{An example of a vehicle platoon modeled by a directed path graph.}
    \label{fig_path_graph}
    \vspace*{-1em}
\end{figure}

In this paper, we consider the multi-agent system that can be modeled by a (directed) linear graph $\mathcal{G}_L$ (also known as a path graph, see \cite{wiki_line_graph})  consisting $K + 1$ agents. Moreover, $\mathcal{G}_L$ satisfies: the neighbors of the agent $A_i$ are $A_{i-1}$ and $A_{i+1}$ for $i = 1,2,3,\hdots,K-1$, the last agent $A_K$ has only one neighbor $A_{K-1}$, and the leader $A_0$ has no neighbors. This is consistent with a group of vehicles marching in a platoon leaded by the first vehicle. An example of a path graph is given by Fig. \ref{fig_path_graph}. We assume that the leader satisfies:

\textit{Assumption III}: the outputs of the leader $A_0$ at time $t+T$ can be predicted by $\Tilde{y}_0(t+T)$ satisfying Eqn. (\ref{eqn_predictor}) and Assumption II. Furthermore, the derivative of the prediction is bounded by:
\begin{equation}
    \Vert \frac{dg_0}{dt}(x_0(t), u_0(t)) \Vert \leq \sigma,
\end{equation}
where $\sigma \in [0, +\infty)$. \hfill $\Box$

If the output of the leader is defined by an external reference signal $r(t), t\in[0,+\infty)$, then Assumption III indicates: 1) the reference signal $r(t)$ is continuously differentiable and the signal in the future $r(t+T)$ is known; 2) the time derivative $\dot{r}(t)$ of $r(\cdot)$ is bounded by $\sigma \in [0, +\infty)$. %

For the multi-agent system under a linear graph topology, if the followers $A_1, A_2, \hdots, A_K$ are $\alpha$-stable (see the Appendix and \cite{niu2023consensus}), then the local consensus error of the agents will be bounded. Denote the trajectory of the system by:
\begin{equation}
\label{eqn_traj_def}
    z\triangleq\begin{bmatrix}
        x_1^\top&x_2^\top&\hdots&x_K^\top&u_1^\top&u_2^\top&\hdots&u_K^\top
    \end{bmatrix}^\top,
\end{equation}
denote the set of all possible system trajectories by $\mathcal{Z}$. Then we have the following:

\textit{Lemma I}: Let the inputs to the agents $A_1, A_2, \hdots, A_K$ be: 
 \begin{equation}
 \label{eqn_leader_follower_controller}
 \begin{split}
     \dot{u}_{i}=-\alpha_{i}\left(\frac{\partial g_{i}}{\partial u_i}(x_{i},u_{i})\right)^{-1}\sum_{j\in N_{i}}\big(g_{i}(x_{i},u_{i})-g_{j}(x_{j},u_{j})\big),
\end{split}
\end{equation}
where $N_i = \{i-1, i+1\}, 1\leq i\leq K-1; N_K = \{K-1\}.$
Assume that the leader $A_0$ satisfies Assumption III, and the followers satisfy Assumption II. If the multi-agent system is $\alpha$-stable over a compact set $\Gamma \subset \mathcal{Z}$, then for every initial condition $z_0 = z(0) \in \Gamma$, the local consensus error of the agent $i \in \{1,2,3,\hdots,K\}$ is bounded by:
\begin{equation}
\label{eqn_local_error}
\lim\sup_{t\rightarrow\infty}\Vert\sum_{j\in N_i}(\tilde{y}_{i}(t+T) - \tilde{y}_{j}(t+T))\Vert^2 \leq \dfrac{\eta}{\alpha_{\min}},
\end{equation}
where $\eta > 0$ is a constant,  $\alpha_{\min} = \min\{\alpha_1, \alpha_2, \hdots, \alpha_K\}$. \hfill$\Box$

Note that Lemma I gives the local consensus error instead of the leader-follower consensus error. However, thanks to the linear graph topology, the leader-follower consensus error can be eliminated by enlarging the controller speedup $\alpha_{\min}$.

\textit{Proposition I}: Under the (directed) linear graph topology $\mathcal{G}_L$, Assumption II for the followers, Assumption III for the leader, and the assumption of $\alpha$-stability of the system, the controller (\ref{eqn_leader_follower_controller}) achieves leader-follower consensus for initial condition $z(0) \in \Gamma$ as $\alpha_{\min}$ goes to infinity:
\begin{equation}
    \lim_{\alpha_{\min}\rightarrow \infty}\lim\sup_{t\rightarrow \infty} \Vert \tilde{y}_i(t+T) - \tilde{y}_0(t+T)\Vert = 0.
\end{equation}
Please see the Appendix for the proofs of Lemma I and Proposition I. Again, the consensus under control law (\ref{eqn_leader_follower_controller}) is achieved over the agents' predicted outputs instead of the actual outputs, which calls for accurate predictors $g_i(\cdot)$ to reduce the consensus error defined by (\ref{eqn_leaderfollower_require}).

\subsection{Consensus control with Integral Control Barrier Function}
The original consensus controller \cite{niu2023consensus} has the risk of inter-agent collisions. To avoid such a situation, we apply the Integral Control Barrier Function to enforce a safety distance between two neighboring agents. With the I-CBF, the inputs of the $i$-th agent, $i = 1,2,3,\hdots,K$, become:
\begin{equation}
\label{eqn_consensus_input_icbf}
\begin{split}
    \dot{u}_{i}=-\alpha_{i}\left(\frac{\partial g_{i}}{\partial u_i}(x_{i},u_{i})\right)^{-1}\sum_{j\in N_{i}}\big(g_{i}(x_{i},u_{i})\\-g_{j}(x_{j},u_{j})\big) + w_i.
\end{split}
\end{equation}
According to \cite{vogel2003comparison}, it is recommended that the time headway when driving on the road should be at least $2$ seconds. Hence, the distance $D$ between two vehicles should satisfy:
\begin{equation}
\label{eqn_min_dist}
  D \geq k_vV,
\end{equation}
where $k_v \geq 2s$. Since the agents are moving in a 2-D plane, the unsafe areas for agent $A_i$ are circles with radius $k_vV_i$ (where $V_i$ is the speed of the bicycle $i$), and the centers of the circles are the neighbor's location $({z}_{j,1}, z_{j,2})^\top$. This leads to the definition of the safety set for $A_i$:
\begin{equation}
\label{eqn_safety_set}
\begin{split}
    &\mathcal{S}_i:= \{(z_{i,1}, z_{i,2})^\top \in \mathbb{R}^2 :\\ &(z_{i,1} - {z}_{j,1})^2 + (z_{i,2} - {z}_{j,2})^2 \geq k_v^2V_{i}^2, \forall j \in N_i\}.
\end{split}
\end{equation}

To guarantee $\mathcal{S}_i$ is forward-invariant and exponentially-stable, define the barrier function between agent $i$ and $j$ as:
\begin{equation}
    \label{eqn_cbf_agenti}
    h_{i,j}^{(1)}(x_i; x_j) = -k_v^2V_i^2 + (z_{i,1} - {z}_{j,1})^2 + (z_{i,2} - {z}_{j,2})^2.
\end{equation}
Taking the derivative of Eqn. (\ref{eqn_cbf_agenti}) yields:
\begin{equation}
\begin{split}
    \dot{h}^{(1)}_{i,j}(x_i, u_i; x_j) &= -2k_v^2V_ia_i \\&+ 2(z_{i,1} - {z}_{j,1})V_i\cos(\psi_i + \gamma_i)\\& + 2(z_{i,2} - {z}_{j,2})V_i\sin(\psi_i + \gamma_i).
\end{split}
\end{equation}
Speeds $(\dot{z}_{j,1}, \dot{z}_{j,2})^\top$ are taken as constants, because if not, then the safety distance will be zero when the neighbors are moving at the same velocity, which is obviously against our common sense. This setting could also help prevent unforeseen accidents where the predecessor stops in a sudden (e.g., in a traffic pile-up). 
We choose the class-$\mathcal{K}$ function as:
\begin{equation}
    \kappa(h^{(1)}_{i,j}(\cdot)) = h^{(1)}_{i,j}(\cdot).
\end{equation}
Note that the bias term $w_i$ does not appear in the term $\dot{h}^{(1)}_{i,j}(\cdot)$. Therefore, the second order I-CBF must be applied. Define:
\begin{equation}
    h^{(2)}_{i,j}(x_i, u_i; x_j) = \dot{h}^{(1)}_{i,j}(x_i, u_i; x_j) + h^{(1)}_{i,j}(x_i, u_i; x_j),
\end{equation}
then:
\begin{equation}
\begin{split}
    \dot{h}^{(2)}_{i,j}(x_i, u_i; x_j) = \dfrac{\partial h^{(2)}_{i,j}}{\partial x_i}\dot{x_i} + \dfrac{\partial h^{(2)}_{i,j}}{\partial u_i}\dot{u}_i, 
\end{split}
\end{equation}
where,
\begin{equation}
    \dfrac{\partial h^{(2)}_{i,j}}{\partial x_i} = \begin{bmatrix}
        2V_i\cos(\psi_i + \gamma_i) + 2(z_{i,1} - z_{j,1})\\2V_i\sin(\psi_i +\gamma_i) + 2(z_{i,2} - z_{j,2})\\-2k_v^2a_i - 2k_i^2V_i\\\left(\makecell{-2(z_{i,1} - z_{i,2})V_i\sin(\psi + \gamma_i) \\+ 2(z_{i,2} - z_{j,2})V_i\cos(\psi_i + \gamma_i)}\right)
    \end{bmatrix}^\top,
\end{equation}

\begin{equation}
    \dfrac{\partial h^{(2)}_{i,j}}{\partial u_i} = \begin{bmatrix}-2k_v^2V_i\\\left(\makecell{-2(z_{i,1} - z_{j,1})V_i\sin(\psi_i + \gamma_i) \\+ 2(z_{i,2} - z_{j,2})V_i\cos(\psi_i + \gamma_i)}\right)
    \end{bmatrix}^\top,
\end{equation}
and $\dot{u}_i$, $\dot{x}_i$ follows input (\ref{eqn_consensus_input_icbf}) and dynamics (\ref{eqn_bicycle_dynamics}) respectively. The bias term $w_i$ appears as long as the vehicle's velocity $V_i$ is not zero. Therefore, we implicitly assume that the vehicles are always moving during the control period. The Quadratic Program associated with the I-CBF is:
\begin{equation}
\label{eqn_bicycle_qp_original}
\begin{split}
    &\min_{w_i} \Vert w_i\Vert^2\\
    \text{s.t.}\ &\dot{h}^{(2)}_{i,j}(x_i, u_i; x_j) + h^{(2)}_{i,j}(x_i,u_i;x_j) \geq 0, \forall j \in N_i.
\end{split}
\end{equation}
For this particular kinematic bicycle system, we noticed that if the solution of $w_i$ from (\ref{eqn_bicycle_qp_original}) is applied, then the bicycles may diverge excessively from its original trajectory. The reason is because the vehicle system (\ref{eqn_bicycle_dynamics}) is more sensitive to the steering wheel angle $\gamma_i$ instead of the acceleration $a_i$. In other words, a slight change in $\gamma_i$ may result in a huge change on the vehicle's path. To avoid such situation, we add weights to (\ref{eqn_bicycle_qp_original}), yielding a weighted quadratic program:
\begin{equation}
\begin{split}
    &\min_{w_i}\ w_i^\top Qw_i\\
    \text{s.t.}\ &\dot{h}^{(2)}_{i,j}(x_i, u_i; x_j) + h^{(2)}_{i,j}(x_i,u_i;x_j) \geq 0, \forall j \in N_i
\end{split}
\end{equation}
where $Q = \text{diag}(q_1, q_2)$ is a diagonal positive definite matrix, and $q_2 > q_1$. Since the constraints on $h_{i,j}^{(2)}(\cdot)$ didn't change, the safety requirements can still be satisfied. %

\section{Simulation results}
\begin{figure}[h]
    \centering
    \includegraphics[width = 6cm]{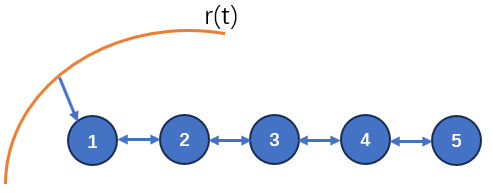}
    \caption{The communication graph of the multi-agent system.}
    \label{fig_comm_graph}
    \vspace*{-1em}
\end{figure}

In this section, we test our proposed consensus controller through a leader-follower consensus control problem consisting six agents. The followers, $A_1, A_2, \hdots, A_5$, are kinematic bicycles following the dynamics of (\ref{eqn_bicycle_dynamics}). The leader, $A_0$, is a virtual agent represented by an external reference signal. The system parameters of the followers are listed in Table \ref{tab_bicycle_parameter}, and their inputs are restricted by $a_i\in[-2,2] (m/s^2)$ and $\gamma_{i} \in [\frac{\pi}{6}, \frac{\pi}{6}] (rad)$ for $i = \{1,2,3,4,5\}$. A path graph is used to describe the agents' communication, as is in Fig. \ref{fig_comm_graph}. 
\begin{table}[h]
    \centering
    \begin{tabular}{|c|c|c|c|c|c|}
    \hline
         Agent Number& 1&2&3&4&5 \\\hline 
         $L_f$&1.105&1.2&1.5&1.2&1.3\\\hline
         $L_r$&1.738&1.7&1.3&1.4&1.3\\\hline
    \end{tabular}
    \caption{System parameters of five kinematic bicycles}
    \vspace*{-2em}
    \label{tab_bicycle_parameter}
\end{table}

\begin{figure}[h]
    \centering
    \includegraphics[width = 8cm]{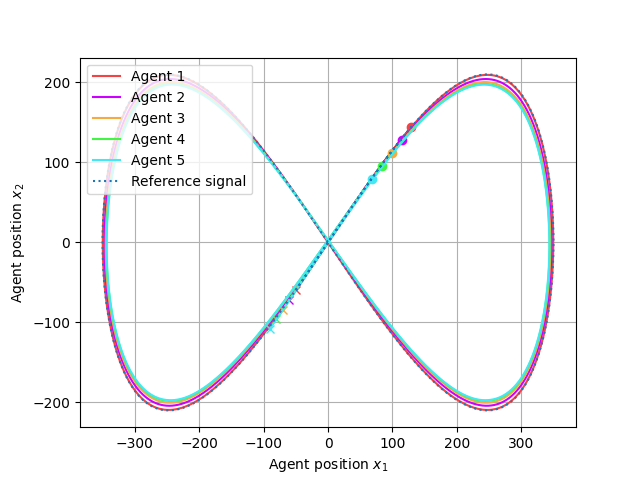}
    \caption{The trajectory of the agents in the bicycle platoon}
    \label{fig_trajectory}
\end{figure}

The reference signal (leader) is specified by:

\begin{equation}
    r(t) = \begin{cases}
    (-50 + 3.75t, -60 + 4.5t)^\top, t \in [0,\frac{40}{3}]\\
   
    \big(350\sin(0.01(t - \frac{40}{3})), 210\sin(0.02(t-\frac{40}{3}))\big)^\top, \\~~~~~~~~~~~~~~~~~~~~~~~~~~~~~t \in [\frac{40}{3}, \infty).
    \end{cases}
\end{equation}
At time $0$, the five agents are arranged in a straight line. The initial locations of the five bicycles are $(-50,-60)$, $(-60,-72)$, $(-70, -84)$, $(-80, -96)$ and $(-90, -108)$, respectively. The controller speedup parameters are $\alpha_i = 10$ for $i = 1,2,\hdots,5$, and the lookahead time horizon is $T = 0.3$.%
The Integral Control Barrier Function applied by agent $A_i, i = 2,3,4,5,$ considers both its successor $A_{i+1}$ and predecessor $A_{i-1}$ (if exists). However, the first agent $A_1$, considers only its successor $A_2$, because the leader $A_0$ is virtual and thus cannot be crashed. The minimal distance that the I-CBF attempts to maintain is chosen to be two times of the agents velocity, indicating $k_v = 2s$. To avoid excessive modifications to the bicycle's steering wheel, the weight matrix is designed as $Q = \text{diag}(1, 999)$. The simulation is carried out using the Forward-Euler method, starting from $t_s = 0s$ until $t_f = 680s$. The simulation stepsize is $0.01s$. The output prediction, $g_i(x_i,u_i; t+T)$, and its partial derivative, $\dfrac{\partial g_i}{\partial u_i}(x_i, u_i; t+T)$, are calculated using explicit Runge-Kutta method of order 5(4) (RK45) using the predictor described in \cite{niu2023consensus}. The trajectories of the agents in the 2-D plain are illustrated in Fig. \ref{fig_trajectory}. The initial positions of the vehicles are marked with "x", and the finial positions are marked with dots.

\begin{figure}[h]
    \centering
    \includegraphics[width = 8cm]{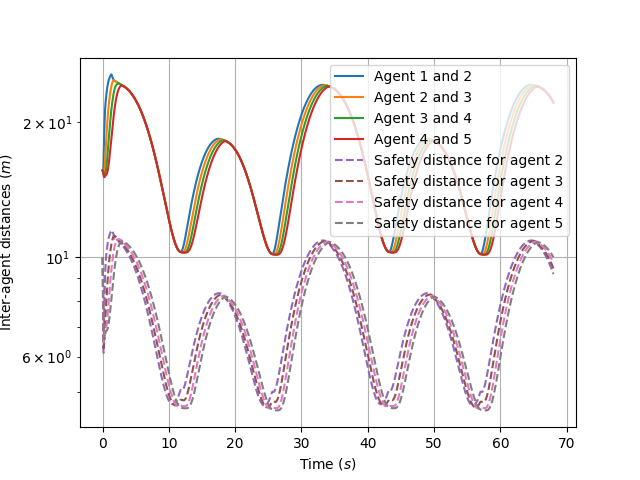}
    \caption{The distance between two adjacency agents in the bicycle platoon}
    \label{fig_agent_distance}
\end{figure}

The distances between two adjacent agents, together with the minimal safety distances defined by (\ref{eqn_min_dist}), are shown in Fig. \ref{fig_agent_distance}. It is shown that the agents' inter-agent distances are always larger than the minimal safety distances, maintaining a headway time of more than two seconds. In real-world driving scenario, this distance allows at least two seconds for the drivers to react if the vehicle in the front stops in a sudden. It is also observed that the inter-agent distances are varying according to the vehicle's velocities, and the minimal and maximal distances appear when the vehicles have the minimal and maximal velocities, respectively. %

\begin{figure}
    \centering
    \includegraphics[width = 8cm]{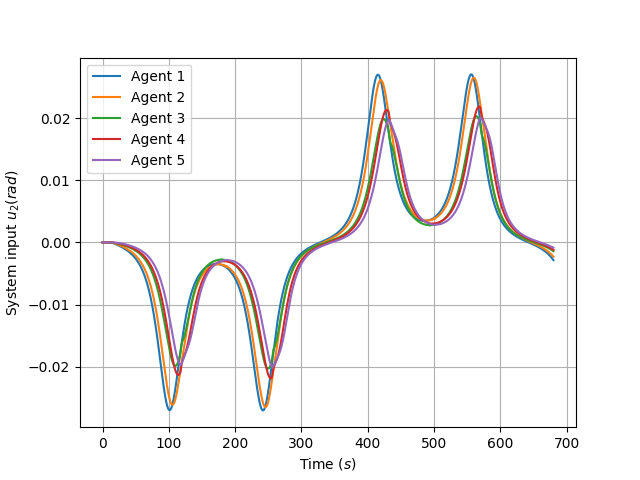}
    \caption{The input $\gamma_i(t)$ of each agent in the platoon}
    \label{fig_agent_u2}
\end{figure}
\begin{figure}[ht]
    \centering
    \includegraphics[width = 8cm]{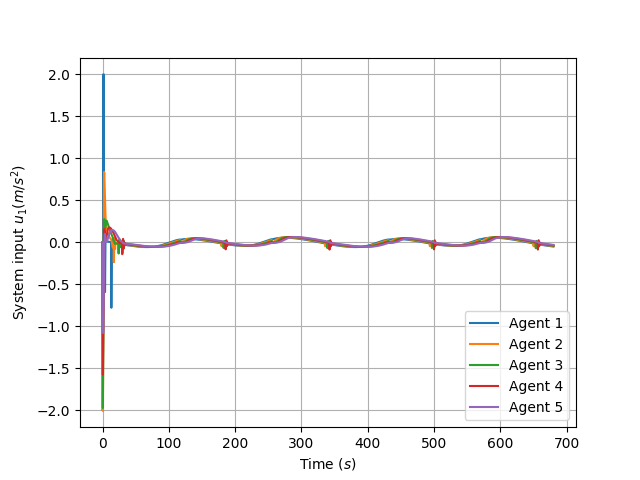}
    \caption{The input $a_i(t)$ of each agent in the platoon}
    \label{fig_agent_u1}
\end{figure}

The agents'  steering angles $\gamma_i(t)$ are shown in Fig. \ref{fig_agent_u2}, while their accelerations $a_i(t)$ are shown in Fig. \ref{fig_agent_u1}. At the beginning of the simulation, the controller exerts large accelerations to reduce inter-agent distances and achieve consensus, leading to initial transients in the inputs. Throughout the simulation, each agent activates the Integral Control Barrier Function for five times. The first activation happens at the beginning when the agents get closed enough, and the following four cases correspond to the four turnings in the vehicles trajectory. Due to the coupling of two neighboring agents, slight input oscillations can be observed in the system inputs when the I-CBF is activated. %



\section{Conclusion}
In this paper, we proposed a consensus controller for multi-agent systems that can also guarantee safety. The consensus control is achieved using output prediction and Newton-Raphson method, and the safety is provided by using the Integral Control Barrier Function. Compared to traditional Control Barrier Functions, the Integral Control Barrier Function is native to the dynamically-defined consensus controller, yielding Quadratic Programs even for nonlinear non-control-affine systems. We applied the proposed consensus controller on a platoon of autonomous vehicles. The vehicles are described by nonlinear kinematic bicycles, and the communication topology is modeled by a path graph. For this special structure, we prove the asymptotic convergence of the leader-follower consensus. Simulation results show that the vehicles can maintain a suitable safety distance while keeping consensus. This approach may have a potential in the control of autonomous vehicles and other similar multi-agent systems.

\section*{Appendix}
\subsection{Proof of Lemma I}
We first state the definition of $\alpha$-stability \cite{niu2023consensus}. In the remainder of this appendix, $z(t)$ represents the trajectory (\ref{eqn_traj_def}) of the multi-agent system, and $\mathcal{Z}$ represents all possible system trajectories. Define $\alpha_{\min} \triangleq \min \{\alpha_1, \alpha_2, \hdots, \alpha_K\}$.%

\textit{Definition I}: The multi-agent system (\ref{eqn_mas_dynamics}) with inputs (\ref{eqn_consensus_controller}) is $\alpha$-stable on a closed set $\Gamma\subset\mathcal{Z}$ if there exists $\Bar{\alpha} >0$, and a continuous, monotonous non-decreasing function $\beta : \mathbb{R}^+\rightarrow\mathbb{R}^+$, such that for all $\alpha_{\min} > \Bar{\alpha}$, the system trajectory $z(t), \forall t \in [0,+\infty)$ is well-defined and satisfies:
\begin{enumerate}
    \item the partial Jacobian matrix $\dfrac{\partial g_i}{\partial u_i}(x_i, u_i)$ is non-singular;
    \item For every initial condition $z_0 = z(0) \in \Gamma$, the trajectory $z(t)$ is bounded by:
    \begin{equation*}
  ~~~~~~~~~~~~~~~\sup_{t \in [0,\infty)}\Vert z(t)\Vert \leq \beta(\Vert z_0\Vert).~~~~~~~~~~~~~~~~~~~~\Box
    \end{equation*}
\end{enumerate}

The definition of the system trajectories and the $\alpha$-stability concerns only the agents controlled by (\ref{eqn_consensus_controller}). This excludes the autonomous agent $A_0$ in the leader-follower consensus. The $\alpha$-stability of nonlinear systems has to be verified case-by-case, and a criterion for single-agent linear systems can be find in \cite{Arxiv}. Based on $\alpha$-stability, Lemma I can be proved.%

\textit{Proof of Lemma I}: For $i = 0, 1,2,\hdots,K-1$, let:
\begin{equation}
    V_i(z(t)) = \Vert g_i(x_i, u_i) - g_{i+1}(x_{i+1}, u_{i+1})\Vert^2. 
\end{equation}
Each $V_i(z(t))$ measures the difference between two predictions of two neighboring agents. Let:
\begin{equation}
    V(z(t)) = \dfrac{1}{2}\sum_{i = 0}^{K-1}V_i(z(t)).
\end{equation}
 Taking the derivative of $V$ yields:
\begin{equation}
\begin{split}
    &\dot{V}(z(t)) = \dfrac{1}{2}\sum_{i = 0}^{K-1}\dot{V}_i(z(t)),
\end{split}
\end{equation}
where,
\begin{equation}
\begin{split}
    \dot{V}_i(z(t)) = \langle &g_i(x_i, u_i)
    - g_{i+1}(x_{i+1}, u_{i+1}),\\& \dot{g}_i(x_i, u_i) - \dot{g}_{i+1}(x_{i+1}, u_{i+1})\rangle.
\end{split}
\end{equation}
Gathering terms of two neighboring terms yields:
\begin{equation}
\begin{split}
&\dot{V}(z(t)) = \langle g_0(x_0, u_0) - g_1(x_1, u_1), \dot{g}_0(x_0, u_0)\rangle+  \\
    &\sum_{i=1}^{K-1}\langle (g_i(x_i, u_i) - g_{i-1}(x_{i-1},u_{i-1}))+ \\&~~~~~~(g_{i}(x_i, u_i) - g_{i+1}(x_{i+1}, u_{i+1})), \dot{g}_i(x_i, u_i)\rangle+\\
    &\langle g_{K}(x_K, u_K) - g_{K-1}(x_{K-1},u_{K-1}) , \dot{g}_K(x_K,u_K)\rangle.
\end{split}
\end{equation}
For $i = 1,2,3, \hdots, K-1$: 
\begin{equation}
\begin{split}
    \dot{g}_i(x_i, u_i) = \dfrac{\partial g_i}{\partial x_i}\dot{x}_i + \dfrac{\partial g_i}{\partial u_i}\dot{u}_i\\ = \dfrac{\partial g_i}{\partial x_i}\dot{x}_i -\alpha_i((g_i(x_i, u_i) - g_{i-1}(x_{i-1},u_{i-1})) \\ +(g_{i}(x_i, u_i) - g_{i+1}(x_{i+1}, u_{i+1}))).
\end{split}
\end{equation}
and for $i = K$:
\begin{equation}
\begin{split}
    \dot{g}_K(x_K, u_K) = \dfrac{\partial g_K}{\partial x_K}\dot{x}_K -\alpha_K(g_K(x_K, u_K) \\- g_{K-1}(x_{K-1},u_{K-1})).
\end{split}
\end{equation}
Therefore, 
\begin{equation}
\begin{split}
        &\dot{V}(z(t)) = \nu_0(z(t)) + \sum_{i = 1}^{K-1} \nu_i(z(t)) + \nu_{K}(z(t)) -\\& \sum_{i = 1}^{K-1}\alpha_{i}\Vert (g_i(x_i, u_i) - g_{i-1}(x_{i-1}, u_{i-1}))\\&~~~~~~+(g_i(x_i, u_i) - g_{i+1}(x_{i+1}, u_{i+1}))\Vert^2 \\&~~ 
         - \alpha_K\Vert g_K(x_K, u_K) - g_{K-1}(x_{K-1}, u_{K-1})\Vert^2,
\end{split}
\end{equation}
where,
\begin{equation}
\begin{split}
    \nu_0(z(t)) &= \langle g_0(x_0, u_0) - g_1(x_1, u_1), \dot{g}_0(x_0, u_0)\rangle\\
 &\leq \Vert g_0(x_0, u_0) - g_1(x_1, u_1)\Vert \Vert \dot{g}_0(x_0, u_0)\Vert, 
\end{split}
\end{equation}
\begin{equation}
    \begin{split}
        &\nu_i(z(t)) = \langle (g_{i}(x_i, u_i) - g_{i-1}(x_{i-1},u_{i-1})) \\&~(g_i(x_i,u_i) - g_{i+1}(x_{i+1}, u_{i+1})),\dfrac{\partial g_i}{\partial x_i}\dot{x}_i\rangle \\&\leq\Vert (g_i(x_i,u_i) - g_{i-1}(x_{i-1},u_{i-1})) \\&~~ + (g_i(x_i,u_i) - g_{i+1}(x_{i+1},u_{i+1}))\Vert\Vert \dfrac{\partial g_i}{\partial x_i}\dot{x}_i\Vert,\\
        &~~~~~i = 1,2,3,\hdots, K-1,
    \end{split}
\end{equation}
\begin{equation}
\begin{split}
    \nu_K(z(t)) = \langle g_K(x_K,u_K) - g_{K-1}(x_{K-1}, u_{K-1}), \dfrac{\partial g_K}{\partial x_{K}}\dot{x}_K\rangle\\
    \leq \Vert g_K(x_K,u_K) - g_{K-1}(x_{K-1}, u_{K-1})\Vert\Vert\dfrac{\partial g_K}{\partial x_K}\dot{x_K}\Vert.    
\end{split}
\end{equation}
By Assumption II, Assumption III, and the assumed $\alpha$-stability, there exists $\eta > 0$, such that:
\begin{equation}
    \sum_{i=0}^{K}\nu_{i}(z(t)) \leq \eta.
\end{equation}
Therefore,
\begin{equation}
\begin{split}
    &\dot{V}(z(t)) \leq \eta -   \sum_{i = 1}^{K-1}\alpha_{i}\Vert (g_i(x_i, u_i) - g_{i-1}(x_{i-1}, u_{i-1}))+ \\&~~~~~~~~~~~~~~~~~~~~~~~~~(g_i(x_i, u_i)- g_{i+1}(x_{i+1}, u_{i+1}))\Vert^2 -\\& ~~~~~~~~~~~~~~\alpha_K\Vert g_K(x_K, u_K) - g_{K-1}(x_{K-1}, u_{K-1})\Vert^2.
\end{split}
\end{equation}
By dropping extra negative terms, we have:
\begin{equation}
\begin{split}
    \dot{V}(z(t)) \leq \eta - \alpha_i\Vert &(g_i(x_i, u_i) - g_{i-1}(x_{i-1}, u_{i-1}))+ \\&(g_i(x_i, u_i)- g_{i+1}(x_{i+1}, u_{i+1}))\Vert^2,\\&i = 1,2,3,\hdots,K-1;
\end{split}
\end{equation}
\begin{equation}
    \dot{V}(z(t)) \leq \eta - \alpha_K\Vert g_K(x_K, u_K) - g_{K-1}(x_{K-1}, u_{K-1})\Vert^2.
\end{equation}
Since $V(z(t))$ is positive definite, by the standard Lyapunov method, the local consensus error satisfies:
\begin{equation}
    \lim\sup_{t\rightarrow\infty}(\eta - \alpha_{\min}\Vert \sum_{j \in N_i}(g_i(x_i,u_i) - g_j(x_j,u_j))\Vert^2) \geq 0,
\end{equation}which indicates Eqn. (\ref{eqn_local_error}). $\blacksquare$

\subsection{Proof of Proposition I}
The proof of Proposition I requires the concept of rooted out-branching (\cite{mesbahi2010graph}, Definition 3.7).

\textit{Definition II}: A directed graph is a rooted out-branching if:
\begin{enumerate}
    \item it does not contain a directed circle, and
    \item it has a vertex $V_r$, such that for every other vertex $V_i, r\not=i$, there exists a directed path from $V_r$ to $V_i$.
\end{enumerate}
Denote the adjacency matrix of the communication graph $\mathcal{G}_L$ by $A \in \mathbb{R}^{(K+1)\times(K+1)}$, such that:
\begin{equation}
    A_{i,j} = \begin{cases}
        1, \text{directed edge $(V_{j-1}, V_{i-1})$ exists}\\0, \text{otherwise.}
    \end{cases}
\end{equation}
The index of the vertex is $j-1,i-1$ because the agents are numbered from $0$ instead of $1$. The directed edge $(V_{j-1}, V_{i-1})$ exists if and only if $A_{i-1}$ can receive information from $A_{j-1}$. Denote the degree matrix of $\mathcal{G}$ by $D$, such that:
\begin{equation}
\begin{split}
    D &= \text{diag}(d_1, d_2, d_3, \hdots, d_{K+1}),\
    d_i = \sum_{j = 1}^{K+1}A_{i,j}.
\end{split}
\end{equation}
Denote the Laplacian of the graph $\mathcal{G}_L$ by $L \triangleq D - A$. Then the following results holds (\cite{mesbahi2010graph}, Proposition 3.8):

\textit{Lemma II}: The directed graph $\mathcal{G}_L$ with $K+1$ nodes contains a rooted out-branching as a subgraph if and only if $\text{rank}(L) = K$. Moreover, if a rooted out-branching exists, then $\text{null}(L) = \text{span}(\mathbf{1}) \triangleq \text{span}([1,1,1,\hdots,1]^\top)$. \hfill$\Box$%

\textit{Proof of Proposition I}: By Eqn. (\ref{eqn_local_error}), since $\eta \not = \infty $, the local consensus error satisfies:
\begin{equation}
    \label{eqn_local_error_inf}
    \lim_{\alpha_{\min}\rightarrow\infty}\lim \sup_{t\rightarrow\infty} \Vert \sum_{j\in N_i} (\tilde{y}_i(t+T) - \tilde{y}_j(t+T))\Vert^2 = 0.
\end{equation}
Define $\tilde{Y} \triangleq [\tilde{y}_0(t+T)^\top,\tilde{y}_1(t+T)^\top,\hdots, \tilde{y}_n(t+T)^\top]^\top$, denote $I_{K+1}$ as the $(K+1)\times (K+1)$ identity matrix. Denote $\otimes$ as the Kronecker product.
Since Eqn. (\ref{eqn_local_error_inf}) holds for all $i \in \{1,2,3,\hdots, K\}$, it follows:
\begin{equation}    \lim_{\alpha_{\min}\rightarrow\infty}\lim\sup_{t\rightarrow\infty} (L\otimes I_{K+1})\tilde{Y} = 0,
\end{equation}
which is:
\begin{equation}
\label{eqn_y_null_space}
    \tilde{Y} \in \text{null}(L\otimes I_{K+1}).
\end{equation}
Observe that the linear graph $\mathcal{G}_L$ contains a rooted out-branching $V_0, V_1, \hdots, V_K$. By Lemma II and the properties of the Kronecker product, the null space of $L\otimes I_{K+1}$ is:
\begin{equation}
    \text{span}(\mathbf{1}\otimes e_1, \mathbf{1}\otimes e_2, \hdots, \mathbf{1}\otimes e_{K+1}),
\end{equation}
where $e_i$ is the unit vector with the $i$-th element being $1$ and others being $0$. Therefore, Eqn. (\ref{eqn_y_null_space}) indicates:
\begin{equation}
    \tilde{y}_0(t+T) = \tilde{y}_1(t+T) = \hdots = \tilde{y}_n(t+T),
\end{equation}
and thus Proposition I holds. $\blacksquare$

\end{document}